\documentclass[letterpaper, 10 pt, conference]{ieeeconf}  

\IEEEoverridecommandlockouts                              

\usepackage{graphics} 
\usepackage{epsfig} 
\usepackage{booktabs} 
\usepackage{multirow} 
\usepackage{xcolor}
\usepackage{hyperref}
\usepackage{url}
\pdfoutput=1

\title{\LARGE \bf
Blind microscopy image denoising with a deep residual and multiscale encoder/decoder network.}

\author{Fabio Hernán Gil Zuluaga$^{1,2}$, Francesco Bardozzo$^{1}$, Jorge Iván Ríos Patiño$^{2}$, \\ Roberto Tagliaferri$^{1}$,~\IEEEmembership{Senior Member,~IEEE}
\thanks{*This work was supported by Università degli Studi di Salerno}
\thanks{$^{1,2}$ Fabio Hernán Gil Zuluagar is with Faculty of Engineering, Universidad Tecnológica de Pereira and Neuronelab, DISA-MIS, University Degli Study di Salerno.
        {\tt\small fhgil@utp.edu.co}}%
        
\thanks{$^{1}$ Francesco Bardozzo is with Neuronelab, DISA-MIS, University Degli Study di Salerno
        {\tt\small fbardozzo@unisa.it}}%

\thanks{$^{2}$ Jorge Iván Ríos Patiño is with Faculty of Engineering, Universidad Tecnológica de Pereira
        {\tt\small jirios@utp.edu.co}}%

\thanks{$^{1}$ Roberto Tagliaferri is with Neuronelab, DISA-MIS, University Degli Study di Salerno
        {\tt\small robtag@unisa.it}}%
}

\begin{document}

\maketitle
\thispagestyle{empty}
\pagestyle{empty}

\begin{abstract}
In computer-aided diagnosis (CAD) focused on microscopy,
denoising improves the quality of image analysis. In
general, the accuracy of this process may depend both on
the experience of the microscopist and on the equipment
sensitivity and specificity. A medical image could be
corrupted by both intrinsic noise, due to the device
limitations, and, by extrinsic signal perturbations during
image acquisition. Nowadays, CAD deep learning applications
pre-process images with image denoising models to reinforce
learning and prediction. In this work, an innovative and
lightweight deep multiscale convolutional encoder-decoder
neural network is proposed. Specifically, the encoder uses
deterministic mapping to map features into a hidden
representation. Then, the latent representation is rebuilt
to generate the reconstructed denoised image. Residual
learning strategies are used to improve and accelerate the
training process using skip connections in bridging across
convolutional and deconvolutional layers. The proposed
model reaches on average 38.38 of PSNR and 0.98 of SSIM on
a test set of 57458 images overcoming state-of-the-art
models in the same application domain.

\indent \textit{Clinical relevance} - Encoder-decoder based denoiser enables
industry experts to provide more accurate and reliable
medical interpretation and diagnosis in a variety of
fields, from microscopy to surgery, with the benefit of
real-time processing.
\end{abstract}

\begin{figure*}[!h]
    \centering
    \includegraphics[scale=0.55]{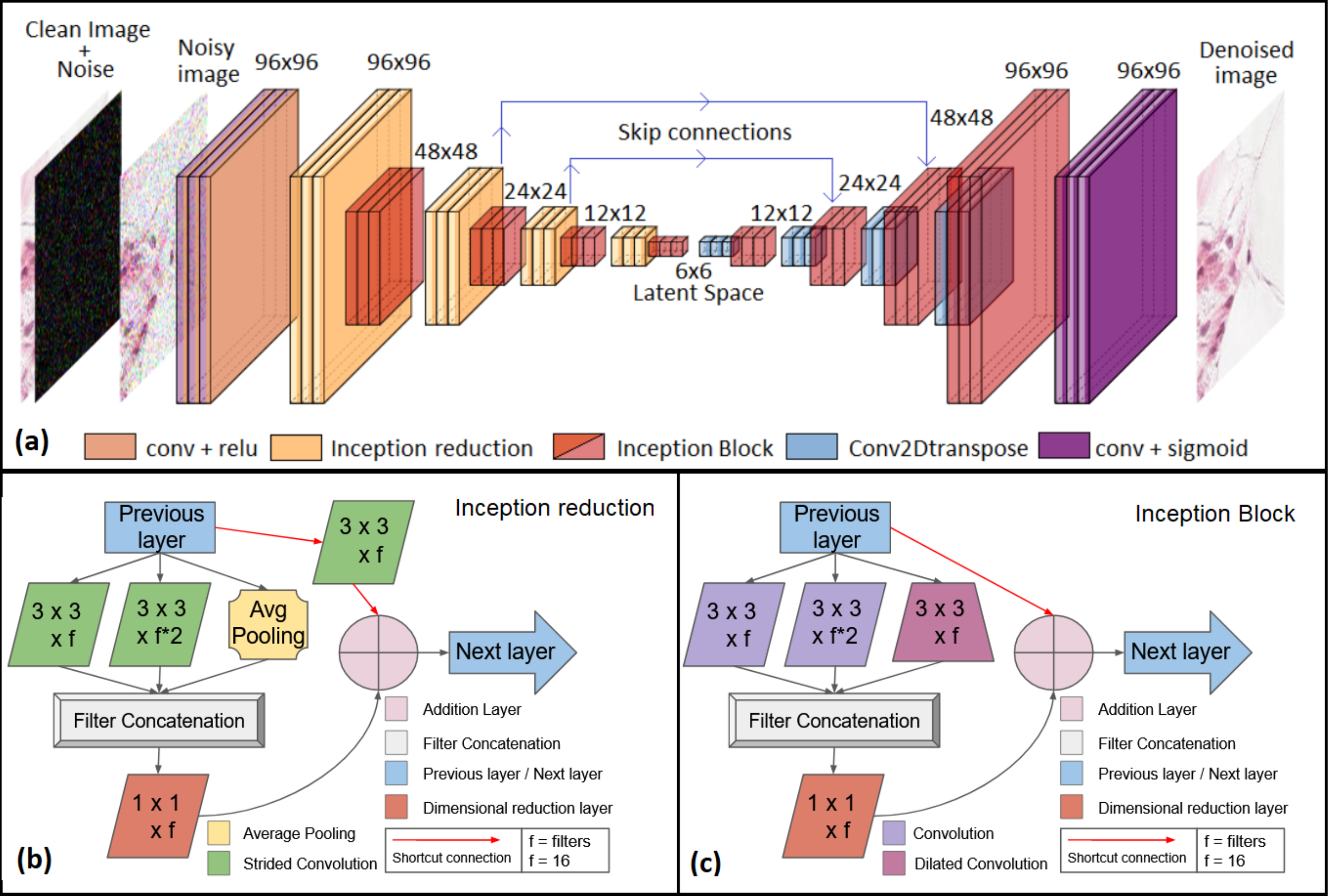}
    \caption{\label{fig1} Figure \ref{fig1} - Box \textbf{(a)} shows the proposed network architecture: the noised image $x^*$ is given as input, the first layer ( conv + relu) maps the initial features followed by four \emph{inception reduction} and \emph{inception blocks} building the latent space $y$. The image reconstruction is composed of four transposed convolutions and inceptions blocks, the last layer is a convolutional layer with a sigmoid activation function (conv + sigmoid). Figure \ref{fig1} Box \textbf{(b)} shows the proposed \emph{inception reduction block}, the main branch has two strided convolution and average pooling that are merged in the concatenation layer. They are followed by a dimensional reduction layer. The previous layer uses the \emph{shortcut connections} for residual learning by executing a strided convolution to match the spatial reduction occurred in the main path due to strided convolutions and average pooling. The addition layer, at the end, sum the weights and passes the output to the next layer. Fig. \ref{fig1} - Box \textbf{(c)} shows the proposed \emph{inception block}: the main path has two convolutions and a dilated convolution that are merged in the concatenation layer. They are followed by a dimensional reduction layer. The previous layer uses the \emph{shortcut connections} for residual learning. Finally, the addition layer sum the weights and passes the output to the next layer.}
    \label{Fig. 1}
\end{figure*}

\section{INTRODUCTION}
Medical image denoising is a well-known ill-posed inverse problem that has been extensively studied in the past decades (traditional models) and recently improved with deep learning approaches. It is possible to divide the traditional models into four main typologies:   \textbf{(i)} \emph{Spatial Domain Filtering} approaches, with Least Mean, Bilateral filtering, Non-Local mean (NLM) and K-Means Singular Value Decomposition (K-SVD) \cite{6587769}. \textbf{(ii)} \emph{Transform Domain Filtering} with Fast Fourier Transform (FFT), Discrete Cosine Transform (DCT), wavelets, curvelets and Block Matching 3-D (BM3D) \cite{lebrun2012secrets},\cite{dabov2007image}. \textbf{(iii)} Other domains covered by \emph{Markov Random Fields} (MRF) \cite{rangarajan1995markov}, Maximum a posteriori probability estimator (MAP) \cite{rabbani2009wavelet}, \textbf{(iv)} \emph{Sparse Representations} with learned simultaneous sparse coding (LSSC) \cite{mairal2009non}, Convolution Sparse Representation (CSR) \cite{yang2014image}. However, for a complete survey please refer to \cite{goyal2020image}. Modern applications for medical image denoising are mainly developed with deep learning models. In \cite{gondara2016medical}, for example,  ad-hoc convolutional denoising autoencoders (CDAE) are used to denoise medical images corrupted with different noise types. Later in \cite{mao2016image}, an encoder-decoder neural network is designed to handle different noise levels by introducing skip connections. In the following year, in \cite{zhang2017beyond}, a very deep convolutional neural network faced the problem of denoising using residual learning \cite{he2016deep} and batch normalization \cite{ioffe2015batch}. Moreover, in \cite{zhang2017learning}, a variable splitting technique is used for denoising. In \cite{zhang2018ffdnet}, a different approach with reversible downsampling operation and tunable noise map is proved to be an effective denoising method. For example,  \cite{jifara2019medical,zhang2017beyond} improve the chest radiographs reconstruction quality with slight modifications from the previous cited models. In \cite{sharif2020learning}, a dynamic residual attention network with noise gate is introduced to denoise medical images of different typologies. With respect to previous models, our work introduces a lightweight convolutional neural network, making possible to transfer the trained networks on Lab-On-Chip applications. Furthermore, the introduced model obtains better results with respect to state-of-the-art models with a relevant generalization power.  In fact, our model is able to deal with unknown noise characteristics (blind denoising) in a wide range of $\sigma$ ($\sigma \in [0,50]$). Moreover, the proposed model is able to significantly reduce artifacts as a result of the implemented multiscale layered architecture (see also Fig \ref{Fig. 1}).  The paper is organized as follows: in section \ref{methods} the model architecture and the dataset are described. In section \ref{res_and_dis} experiments and results are presented and discussed, followed by section \ref{conclusion} with the conclusions.

\section{Methods}
\label{methods}
In section \ref{dataset} the dataset is provided, while in section \ref{preproc} the procedure for preprocessing is shown. Finally, the model architecture is explained in section \ref{mod_arch}. 

\subsection{Dataset}
\label{dataset}
\begin{figure}[]
    \centering
    \includegraphics[scale=0.80]{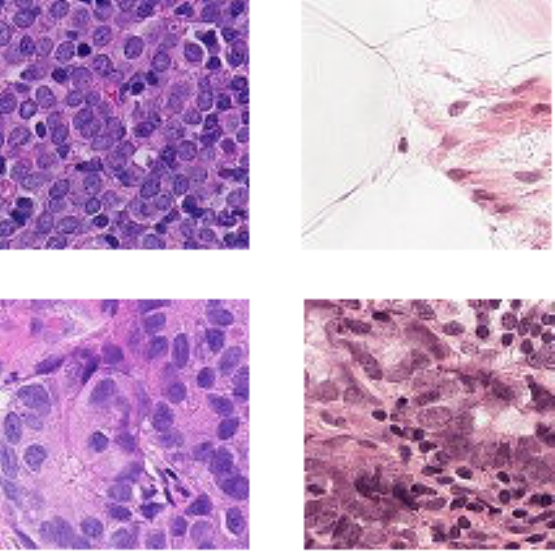}
    \caption{\label{fi2} The figure shows four microscopy image tissues from the Histopathologic Cancer Detection dataset}
\end{figure}

Our deep learning model is trained and tested on a large collection of microscopy images from Histopathologic ection Dataset\footnote{Link: \href{https://www.kaggle.com/c/histopathologic-cancer-detection}{Histopathologic Cancer Detection Dataset }}. In total the dataset contains 220025 training microscopy images and 57458 test images with a size of $96 \times 96$ pixels on three channels (RGB). In detail, Fig. \ref{fi2} shows a sample set, illustrating various degrees of luminance, contrast and structure.

\subsection{Image pre-processing and experimental setup}
\label{preproc}
A synthetic generated Additive white Gaussian noise (AWGN) is added to the microscopy images. AWGN follows the standard assumption that there is no prior information of the type of noise perturbation. This is also according to \cite{lee1981refined} where real-world noise can be approximated as locally AWGN. The noise generator is built with the numpy library \cite{oliphant2006guide}. The $220025$ images were corrupted with standard deviation in the range between 0 and 50 ($\sigma \in [0,50]$). The perturbations are equally distributed over the total of the images, obtaining sets of 4314 images each one belonging to a  $\sigma$ level (e.g 4314 images for $\sigma = 1$, 4314 images for $\sigma = 2$, and so on).  In this way we obtained a training set of images divided into $51$ subsets each one with a different value of $\sigma$ from 0 to 50. The proposed model is trained by using all these subsets. Moreover, despite the work in \cite{zhang2018ffdnet}, \cite{zhang2017learning} and \cite{jifara2019medical}, in which the training dataset was generated with fixed $\sigma$, we used a multi sigma training set; in fact, these works trained different networks for each specific sigma while we trained a single network capable of handling noise levels, ranging from 0 to 50.  In detail, this means that the network is able to perform the so called \emph{blind denoising procedure} after training. In other words, our network can deal with noisy images without knowing its characteristic perturbations (i.e. noise intensity, distribution, standard deviation, etc...).  The different noise levels are generated with a fixed seed to ensure fair comparison and experiment reproducibility.  In detail, all the noise maps are created with mean ($\mu$ = 0), $\sigma \in [0,50]$. It is performed a pixel-wise 8-bit quantization in range $[0, 255]$ (please for more details refer to our online repository \footnote{\href{https://github.com/Fabio-Gil-Z/IRUNet}{GitHub Repository - IRUNet}})

\subsection{Model architecture}
\label{mod_arch}

\begin{figure}[!h]
    \centering
    \includegraphics[scale=0.40]{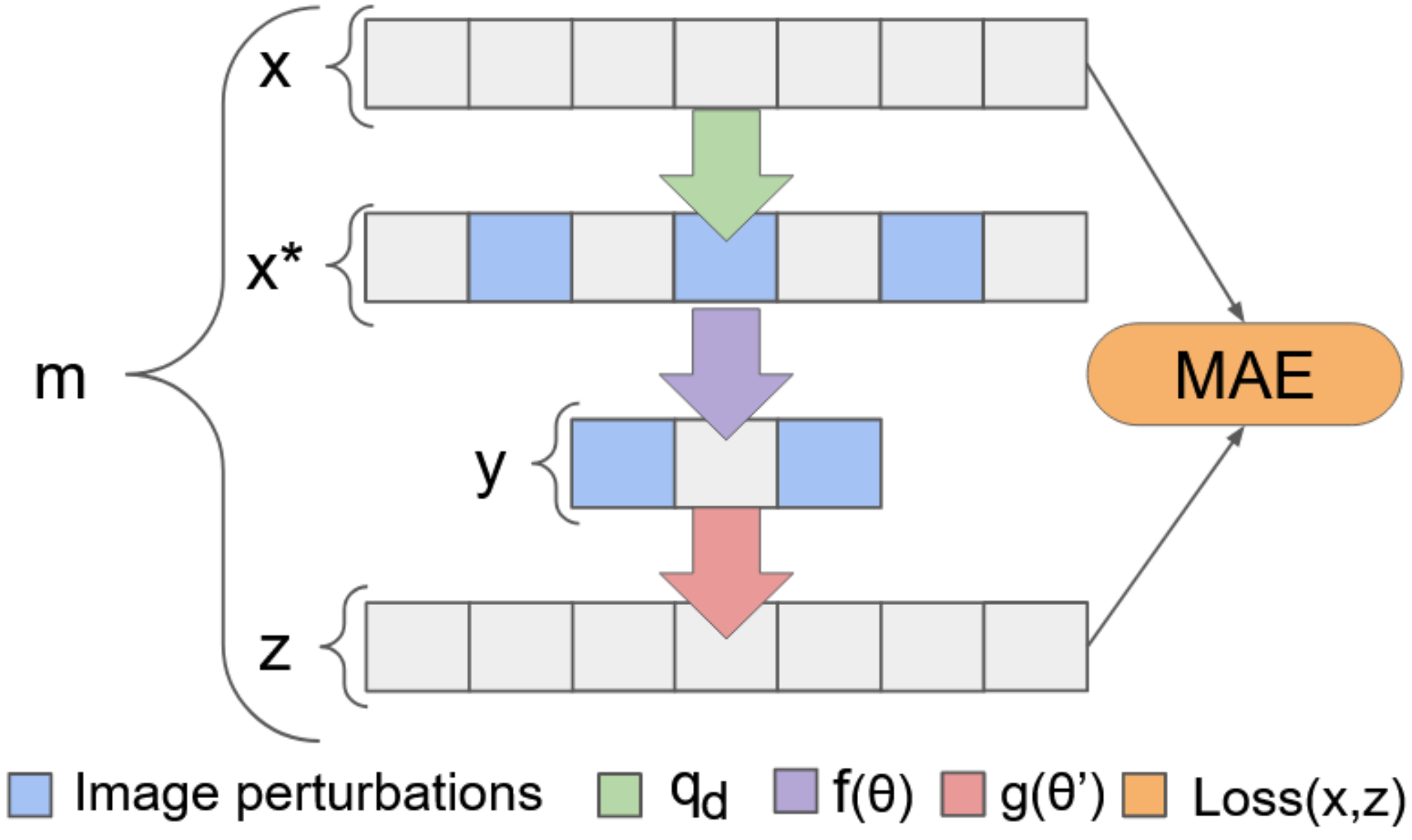}
    \caption{\label{fig3} Encoder-decoder pipeline: the perturbation of the clean image \emph{x} is done by ${q}_d$ obtaining the noisy image ${x}^*$. The encoder ${f}_{\theta}$ maps into a latent space \emph{y}. The decoder ${g}_{\theta '}$ takes the latent space as input and outputs an approximation of \emph{x}, producing \emph{z}. Finally, the model tries to minimize during the epochs the reconstruction error between \emph{x} and \emph{z} (loss(x,z)).}
    \label{Fig. 3}
\end{figure}

Our model architecture is inspired to the unsupervised denoising autoencoders provided by \cite{vincent2008extracting,gondara2016medical}. However, it is not an autoencoder (in the strict sense of the term) but, more precisely, an encoder/decoder network. The whole model architecture is described in Figure \ref{fig1}. Given $x$ the clean image and $x^{*}$ the noised one, the objective is to learn a mapping from ${x}^*$ (noisy image) to its denoised representation $z$ (reconstructed image). Formally, the model $m$ can be represented as $m(x\mid{x}^*;\theta)$; with $\theta$ parameters to be learned. Initially, as it is shown in Figure \ref{fig3}, the original microscopy image can be represented as a \emph{d}-dimensional space with pixel intensities normalized between 0 and 1 ($x \in [0,1{]}^d$). Then, this space is corrupted by means of a stochastic mapping $x^{*} \sim {q}_d({x}^*\mid x)$ where ${x}^*$ is a corrupted version of $x$ (see also Section \ref{preproc}). In the encoder ${f}_\theta$, the corrupted ${x}^*$ is mapped into a hidden representation $ y = {f}_\theta({x}^*) = \delta_{W,b}({x}^*) $. The activation function $\delta$ is the Rectified Linear Unit (ReLU)  \cite{nair2010rectified}. While, the learnable parameter $\theta$ is equal to \{W,b\}, with \emph{W} the weight matrices and \emph{b} the biases. The decoder ${g}_{\theta '}$ reconstruct the original image from the latent space ( $z = {g}_{\theta^{'}}(y) = \delta^{'}_{W^{'},b^{'}}(y)$. The parameters $ W,b,{W}^{'},{b}^{'} $ are obtained by minimizing the reconstruction error between the original image (\emph{x}) and the reconstructed one (\emph{z}) (see also Fig \ref{fig3}). The mean absolute error (MAE) is the loss function that our optimizer tries to minimize. In detail, the model architecture is designed leveraging the interplay between two \emph{inception blocks} \cite{szegedy2017inception} (Figure \ref{fig1} - Box \textbf{(b)} and \textbf{(c)}). According to the denoise model of \cite{song2020grouped}, the multiscale configuration is adopted because it performs better on difficult image microscopy areas (edges and homogeneous textures). To reduce the vanishing gradient problem, the network architecture is designed, in a certain sense, wider rather than deeper \cite{ hochreiter1998vanishing} with a strategic positioning of \emph{skip connections}. In detail, two different types of \emph{skip connections} (by layer concatenation) are designed to provide an alternative gradient path in backpropagation. The first type of skip connections are positioned between the encoder and the decoder (see Figure \ref{fig1} - Box \textbf{(a)}). In detail, they are typically adopted to avoid information loss (see also \cite{mao2016image,ronneberger2015u}). The second type of skip connections are suited inside the two \emph{inception blocks} and named \emph{shortcut connections} (see also Figure \ref{fig1} - Box \textbf{(b)} and \textbf{(c)}). In some situations, \emph{shortcut connections} increase model accuracy by leveraging residual learning approaches \cite{zhang2017beyond,zhang2017learning}

\section{Experiments and Discussion}
\label{res_and_dis}
In section \ref{trainingtime}, training process and prediction time are presented. In section \ref{res} our model results are shown in comparison with traditional \cite{dabov2007image} and state-of-the-art deep learning models \cite{zhang2017beyond,jifara2019medical,sharif2020learning}.

\subsection{Model configurations}
\label{trainingtime}
The network is trained with Adam \cite{kingma2014adam} optimizer for a total of $123.379$ trainable parameters with $b_1=0.9$, $b_2=0.999$ and $\epsilon$ equal to $1*10^{-7}$. The learning rate is of $1*10^{-4}$. The hyperparameters tuning comes through a grid search on filter selection, learning rate monitoring, skip connection positioning and several cost functions testing. The evaluation of predictions and model performances are based on  PSNR evaluations. The model was trained with  Nvidia GeForce GTX 1080, processor Intel® Xeon(R) CPU E5-2630 v4 @ 2.20GHz × 20, employing Tensorflow v2.2, Cuda and Cudnn v10.1 with Python v3.8. The whole training process took about three days. Regarding computation performance, the average prediction time of the network is approximately 0.03 seconds per image.

\subsection{Model performances}
\label{res}
\begin{figure}[!h]
    \centering
    \includegraphics[scale=0.90]{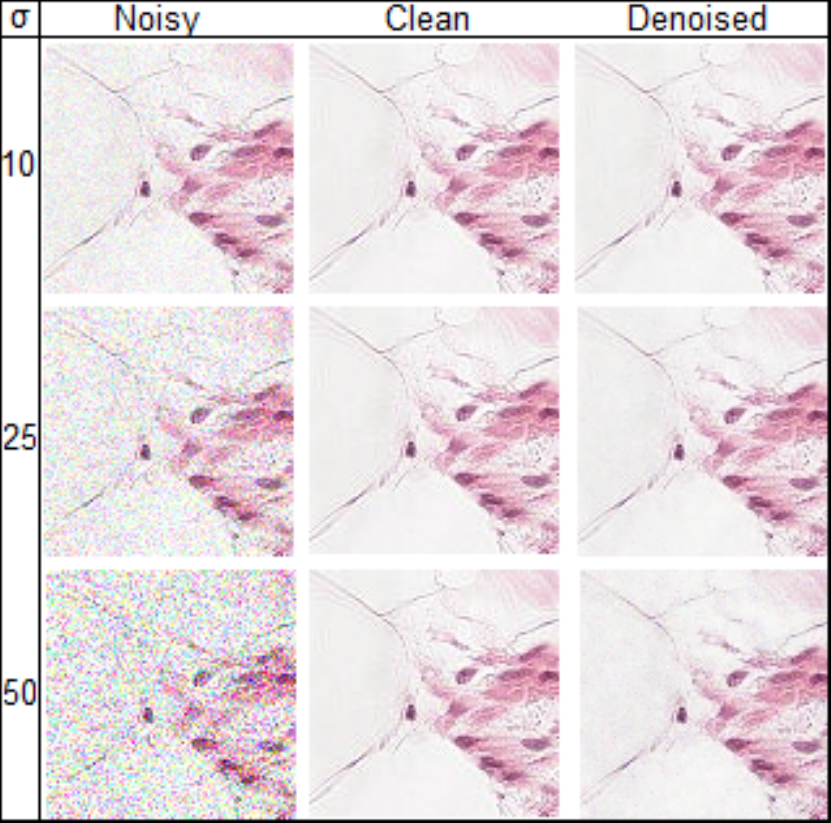}
    \caption{\label{fig4} In Figure \ref{fig4} three samples perturbed by three $\sigma = [10,25,50]$ noise levels are shown.  In detail, in the first coloumn (\emph{Noisy}) the noisy images are depicted, while the ground truth is labelled as \emph{Clean}, finally, in the third coloumn, the denoised images are shown. As it is described in Section \ref{res}, our model is able to remove the various levels of noise over the homogeneous areas and along the edges.}
    \label{Fig. 4}
\end{figure}

As it is shown in Table \ref{table1}, the proposed architecture outperforms the other denoising methods using as reconstruction measures PSNR and SSIM; for PSNR at $\sigma = 10$ the difference between the proposed method and the second and third top methods are $2.84 dB$ and $5.27 dB$, respectively; at $\sigma = 25$ the gap increases to $9.66 dB$ with respect to DRAN and $10.41 dB$ to Residual MID. When $\sigma = 25$ it is obtained the biggest improvement over the three $\sigma$ evaluations. In the last comparison, with $\sigma = 50$, the differences  with the second and third best evaluations were $5.25 dB$ and $11.66 dB$, respectively. Similar results can be seen for SSIM, when our network reached the highest values in all three $\sigma$ evaluations, being the only network with values over $0.96$. In Fig. \ref{fig4}, the reconstruction quality can be evaluated considering homogeneous areas, edges and image borders.

\begin{table}[!h]
\centering
\caption{\label{table1} Results and comparisons}
\begin{tabular}{|l|c|r|r|}
\hline
Model & \multicolumn{1}{l|}{$\sigma$} & \multicolumn{1}{l|}{PSNR} & \multicolumn{1}{l|}{SSIM} \\ \hline
BMD3* \cite{dabov2007image}                        & \multirow{5}{*}{10} & 28.19          & 0.6670          \\ \cline{1-1} \cline{3-4} 
DnCNN \cite{zhang2017beyond}                      &                     & 35.26          & 0.8119          \\ \cline{1-1} \cline{3-4} 
Residual MID \cite{jifara2019medical}               &                     & 36.93          & 0.8769          \\ \cline{1-1} \cline{3-4} 
DRAN \cite{sharif2020learning}                       &                     & 39.36          & 0.9735          \\ \cline{1-1} \cline{3-4} 
\textbf{IRUNet (Proposed)} &                     & \textbf{42.20} & \textbf{0.9977} \\ \hline
BMD3* \cite{dabov2007image}                        & \multirow{5}{*}{25} & 25.02          & 0.5042          \\ \cline{1-1} \cline{3-4} 
DnCNN \cite{zhang2017beyond}                      &                     & 26.70          & 0.7976          \\ \cline{1-1} \cline{3-4} 
Residual MID \cite{jifara2019medical}               &                     & 29.23          & 0.8518          \\ \cline{1-1} \cline{3-4} 
DRAN \cite{sharif2020learning}                       &                     & 29.98          & 0.8993          \\ \cline{1-1} \cline{3-4} 
\textbf{IRUNet (Proposed)} &                     & \textbf{39.64} & \textbf{0.9925} \\ \hline
BMD3* \cite{dabov2007image}                        & \multirow{5}{*}{50} & 20.14          & 0.4248          \\ \cline{1-1} \cline{3-4} 
DnCNN \cite{zhang2017beyond}                      &                     & 21.49          & 0.5046          \\ \cline{1-1} \cline{3-4} 
Residual MID \cite{jifara2019medical}               &                     & 21.65          & 0.5652          \\ \cline{1-1} \cline{3-4} 
DRAN \cite{sharif2020learning}                       &                     & 28.06          & 0.8198          \\ \cline{1-1} \cline{3-4} 
\textbf{IRUNet (Proposed)} &                     & \textbf{33.31} & \textbf{0.9655} \\ \hline
\end{tabular}

\begin{itemize}
    \centering
    \item Note: The traditional model is shown with \textbf{*}.
\end{itemize}
\end{table}

\section{Conclusion}
 \label{conclusion} We presented a novel light weight CNN, that compares well with state-of-the-art methodologies both classical and deep neural networks. Our model takes advantages both from its architecture and from the learning of multi-$\sigma$ images. Given the reduced number of learned parameters, the trained network can work on Lab-On-Chip applications. Future work includes new medical image typologies and higher degrees of noise map spatial distributions to increase the generalization power.

\addtolength{\textheight}{-12cm}   







\bibliographystyle{IEEEtran}
\bibliography{bibliography.bib}

\begin{thebibliography}{10}
\providecommand{\url}[1]{#1}
\csname url@samestyle\endcsname
\providecommand{\newblock}{\relax}
\providecommand{\bibinfo}[2]{#2}
\providecommand{\BIBentrySTDinterwordspacing}{\spaceskip=0pt\relax}
\providecommand{\BIBentryALTinterwordstretchfactor}{4}
\providecommand{\BIBentryALTinterwordspacing}{\spaceskip=\fontdimen2\font plus
\BIBentryALTinterwordstretchfactor\fontdimen3\font minus
  \fontdimen4\font\relax}
\providecommand{\BIBforeignlanguage}[2]{{%
\expandafter\ifx\csname l@#1\endcsname\relax
\typeout{** WARNING: IEEEtran.bst: No hyphenation pattern has been}%
\typeout{** loaded for the language `#1'. Using the pattern for}%
\typeout{** the default language instead.}%
\else
\language=\csname l@#1\endcsname
\fi
#2}}
\providecommand{\BIBdecl}{\relax}
\BIBdecl

\bibitem{6587769}
L.~Shao, R.~Yan, X.~Li, and Y.~Liu, ``From heuristic optimization to dictionary
  learning: A review and comprehensive comparison of image denoising
  algorithms,'' \emph{IEEE Transactions on Cybernetics}, vol.~44, no.~7, pp.
  1001--1013, 2014.

\bibitem{lebrun2012secrets}
M.~Lebrun, M.~Colom, A.~Buades, and J.-M. Morel, ``Secrets of image denoising
  cuisine,'' \emph{Acta Numerica}, vol.~21, no.~1, pp. 475--576, 2012.

\bibitem{dabov2007image}
K.~Dabov, A.~Foi, V.~Katkovnik, and K.~Egiazarian, ``Image denoising by sparse
  3-d transform-domain collaborative filtering,'' \emph{IEEE Transactions on
  image processing}, vol.~16, no.~8, pp. 2080--2095, 2007.

\bibitem{rangarajan1995markov}
A.~Rangarajan and R.~Chellappa, ``Markov random eld models in image
  processing,",'' \emph{The Handbook of Brain Theory and Neural Networks, MA
  Arbib, ed}, pp. 564--567, 1995.

\bibitem{rabbani2009wavelet}
H.~Rabbani, R.~Nezafat, and S.~Gazor, ``Wavelet-domain medical image denoising
  using bivariate laplacian mixture model,'' \emph{IEEE transactions on
  biomedical engineering}, vol.~56, no.~12, pp. 2826--2837, 2009.

\bibitem{mairal2009non}
J.~Mairal, F.~Bach, J.~Ponce, G.~Sapiro, and A.~Zisserman, ``Non-local sparse
  models for image restoration,'' in \emph{2009 IEEE 12th international
  conference on computer vision}.\hskip 1em plus 0.5em minus 0.4em\relax IEEE,
  2009, pp. 2272--2279.

\bibitem{yang2014image}
H.-Y. Yang, X.-Y. Wang, P.-P. Niu, and Y.-C. Liu, ``Image denoising using
  nonsubsampled shearlet transform and twin support vector machines,''
  \emph{Neural networks}, vol.~57, pp. 152--165, 2014.

\bibitem{goyal2020image}
B.~Goyal, A.~Dogra, S.~Agrawal, B.~Sohi, and A.~Sharma, ``Image denoising
  review: From classical to state-of-the-art approaches,'' \emph{Information
  fusion}, vol.~55, pp. 220--244, 2020.

\bibitem{gondara2016medical}
L.~Gondara, ``Medical image denoising using convolutional denoising
  autoencoders,'' in \emph{2016 IEEE 16th International Conference on Data
  Mining Workshops (ICDMW)}.\hskip 1em plus 0.5em minus 0.4em\relax IEEE, 2016,
  pp. 241--246.

\bibitem{mao2016image}
X.-J. Mao, C.~Shen, and Y.-B. Yang, ``Image restoration using convolutional
  auto-encoders with symmetric skip connections,'' \emph{arXiv preprint
  arXiv:1606.08921}, 2016.

\bibitem{zhang2017beyond}
K.~Zhang, W.~Zuo, Y.~Chen, D.~Meng, and L.~Zhang, ``Beyond a gaussian denoiser:
  Residual learning of deep cnn for image denoising,'' \emph{IEEE transactions
  on image processing}, vol.~26, no.~7, pp. 3142--3155, 2017.

\bibitem{he2016deep}
K.~He, X.~Zhang, S.~Ren, and J.~Sun, ``Deep residual learning for image
  recognition,'' in \emph{Proceedings of the IEEE conference on computer vision
  and pattern recognition}, 2016, pp. 770--778.

\bibitem{ioffe2015batch}
S.~Ioffe and C.~Szegedy, ``Batch normalization: Accelerating deep network
  training by reducing internal covariate shift,'' in \emph{International
  conference on machine learning}.\hskip 1em plus 0.5em minus 0.4em\relax PMLR,
  2015, pp. 448--456.

\bibitem{zhang2017learning}
K.~Zhang, W.~Zuo, S.~Gu, and L.~Zhang, ``Learning deep cnn denoiser prior for
  image restoration,'' in \emph{Proceedings of the IEEE conference on computer
  vision and pattern recognition}, 2017, pp. 3929--3938.

\bibitem{zhang2018ffdnet}
K.~Zhang, W.~Zuo, and L.~Zhang, ``Ffdnet: Toward a fast and flexible solution
  for cnn-based image denoising,'' \emph{IEEE Transactions on Image
  Processing}, vol.~27, no.~9, pp. 4608--4622, 2018.

\bibitem{jifara2019medical}
W.~Jifara, F.~Jiang, S.~Rho, M.~Cheng, and S.~Liu, ``Medical image denoising
  using convolutional neural network: a residual learning approach,'' \emph{The
  Journal of Supercomputing}, vol.~75, no.~2, pp. 704--718, 2019.

\bibitem{sharif2020learning}
S.~Sharif, R.~A. Naqvi, and M.~Biswas, ``Learning medical image denoising with
  deep dynamic residual attention network,'' \emph{Mathematics}, vol.~8,
  no.~12, p. 2192, 2020.

\bibitem{lee1981refined}
J.-S. Lee, ``Refined filtering of image noise using local statistics,''
  \emph{Computer graphics and image processing}, vol.~15, no.~4, pp. 380--389,
  1981.

\bibitem{oliphant2006guide}
T.~E. Oliphant, \emph{A guide to NumPy}.\hskip 1em plus 0.5em minus 0.4em\relax
  Trelgol Publishing USA, 2006, vol.~1.

\bibitem{vincent2008extracting}
P.~Vincent, H.~Larochelle, Y.~Bengio, and P.-A. Manzagol, ``Extracting and
  composing robust features with denoising autoencoders,'' in \emph{Proceedings
  of the 25th international conference on Machine learning}, 2008, pp.
  1096--1103.

\bibitem{nair2010rectified}
V.~Nair and G.~E. Hinton, ``Rectified linear units improve restricted boltzmann
  machines,'' in \emph{Icml}, 2010.

\bibitem{szegedy2017inception}
C.~Szegedy, S.~Ioffe, V.~Vanhoucke, and A.~Alemi, ``Inception-v4,
  inception-resnet and the impact of residual connections on learning,'' in
  \emph{Proceedings of the AAAI Conference on Artificial Intelligence},
  vol.~31, no.~1, 2017.

\bibitem{song2020grouped}
Y.~Song, Y.~Zhu, and X.~Du, ``Grouped multi-scale network for real-world image
  denoising,'' \emph{IEEE Signal Processing Letters}, vol.~27, pp. 2124--2128,
  2020.

\bibitem{hochreiter1998vanishing}
S.~Hochreiter, ``The vanishing gradient problem during learning recurrent
  neural nets and problem solutions,'' \emph{International Journal of
  Uncertainty, Fuzziness and Knowledge-Based Systems}, vol.~6, no.~02, pp.
  107--116, 1998.

\bibitem{ronneberger2015u}
O.~Ronneberger, P.~Fischer, and T.~Brox, ``U-net: Convolutional networks for
  biomedical image segmentation,'' in \emph{International Conference on Medical
  image computing and computer-assisted intervention}.\hskip 1em plus 0.5em
  minus 0.4em\relax Springer, 2015, pp. 234--241.

\bibitem{kingma2014adam}
D.~P. Kingma and J.~Ba, ``Adam: A method for stochastic optimization,''
  \emph{arXiv preprint arXiv:1412.6980}, 2014.

\end{thebibliography}

\end{document}